\documentclass[aps,superscriptaddress,preprintnumbers,showpacs,twocolumn]{revtex4}
\usepackage{amssymb}
\usepackage{graphicx}
\usepackage{dcolumn}
\usepackage{bm}

\begin{document}

\newcommand*{\PKU}{School of Physics and State Key Laboratory of Nuclear Physics and
Technology, Peking University, Beijing 100871}\affiliation{\PKU}
\newcommand*{\NTU}{Department of Physics and Center for Theoretical Sciences, National Taiwan University, Taipei 10617}\affiliation{\NTU}
\newcommand*{\IPPC}{Institute of Particle Physics and Cosmology, Department of Physics,
Shanghai JiaoTong University, Shanghai 200240}\affiliation{\IPPC}
\newcommand*{\CHEP}{Center for High-Energy
Physics, Peking University,
Beijing 100871}\affiliation{\CHEP}

\title{Black Holes and Photons with Entropic Force}

\author{Xiao-Gang He}\email{hexg@phys.ntu.edu.tw}\affiliation{\PKU}\affiliation{\NTU}\affiliation{\IPPC}\affiliation{\CHEP}
\author{Bo-Qiang Ma}\email{mabq@phy.pku.edu.cn}\affiliation{\PKU}\affiliation{\NTU}\affiliation{\CHEP}

\begin{abstract}
We study entropic force effects on black holes and photons. We find
that application of an entropic analysis restricts the radial change
$\Delta R$ of a black hole of radius $R_{\mathrm{H}}$, due to a test
particle of a Schwarzschild radius $R_{h}$ moving towards the black
hole by $\Delta x$ near black body surface, to be given by a
relation $R_{\mathrm{H}} \Delta R= R_h \Delta x/2$, or ${\Delta R}/{
\lambdabar_M} =  {\Delta x}/{2 \lambdabar_m}$. We suggest a new rule
regarding entropy changes in different dimensions, $\Delta S= 2\pi k
D \Delta l /\lambdabar,$  which unifies Verlinde's conjecture and
the black hole entropy formula. We also propose to extend the
entropic force idea to massless particles such as a photon. We find
that there is an entropic force on a photon of energy $E_\gamma$,
with $F=G M m_{\gamma}/R^2$, and therefore the photon has an
effective gravitational mass $m_\gamma = E_\gamma/c^2$.
\end{abstract}


%\begin{keyword}
%gravity, entropic force, thermodynamics, holographic screen
%\end{keyword}

\maketitle

\noindent
{\bf Introduction}
\\

Newton's laws of motion and gravity provided a unified description
of motions for observable objects on the Earth and in the sky.  One important formula
in the laws of motion is the second law,
\begin{equation}
F=ma,
\label{Fma}
\end{equation}
which states that a point-like particle of mass $m$ acquires an
acceleration $a$ by a force $F$ acting on that particle. Newton's
law of gravity states that two particles of masses $m$ and $M$
experience an attractive force between each other,
\begin{equation}
F=G\frac{mM}{r^2},
\label{gravity}
\end{equation}
where $G$ is a universal gravitational constant called Newton
constant, and $r$ is the distance between the two particles.
Newton's theory only states how the laws work but not
why they work. Therefore the desire to seek for
explanations why these laws work is beyond the scope of classical
mechanism.

There has been a new conjecture~\cite{Verlinde,Padma,Padma2} to
interpret the gravity as an emergent force based on knowledge of
black hole thermodynamics, relativity, and quantum theory. There
have been a number of discussions~\cite{discuss} concerning the
conjecture, with also some applications to
expansion~\cite{expansion}, acceleration~\cite{acceleration}, and
inflation~\cite{inflation} in cosmology. In this work, we further
study some implications of the entropic force. It is found that
since the entropy of a black hole is completely determined by the
surface, the application of entropic force forces the radial change
$\Delta R$ of a black hole of radius $R_{\mathrm{H}}$, due to a test
particle of a Schwarzschild radius $R_h$ moving towards the black
hole by $\Delta x$ near black hole surface, to be given by a
relation $R_{\mathrm{H}} \Delta R= R_h \Delta x/2$, or ${\Delta R}/{
\lambdabar_M} =  {\Delta x}/{2 \lambdabar_m}$, where $\lambdabar_m$
and $\lambdabar_M$ correspond to the Compton wavelengthes of the
test particle and the black hole respectively. We find a consistent
way to unify black hole properties and Verlinde's conjecture. We
also expand the entropic force idea to massless particles such as a
photon. We find that there is an entropic force on this photon, and
therefore the photon has an effective gravitational mass leading to
blue and red shifts, and also deflection for a photon passing by a
massive body.

Several essential ideas of the entropic force are related to the
properties of a back hole. A black hole is a region of space from
which nothing, including light, can escape. Around a black hole
there is a surface or screen which marks the point of no return,
called event horizon. As a black hole exhibits a remarkable tendency
to increase its horizon surface area under any transformation,
Bekenstein~\cite{Bekenstein} conjectured that a black hole entropy
is proportional to the area $A_{\mathrm{BH}}$ of its event horizon
divided by the Planck area $l_{P}^2$, where $l_P=\sqrt{G\hbar /c^3}$
is the Planck length, $\hbar$ is the Planck constant, and $c$ is the
speed of light. The entropy of a black hole is given
by~\cite{Bardeen,Hawking}
\begin{equation}
S_\mathrm{BH}=\frac{k A_\mathrm{BH}}{4 l_{P}^2},
\end{equation}
where $k$ is the Boltzmann constant. The black holes create and emit
particles as if they were black bodies with temperature
$T_\mathrm{BH}$.

The idea that the black hole information can be thought of as
encoded on a boundary to the region preferably a light-like boundary
like the gravitational horizon, is called the holographic
principle~\cite{thooft,susskind}. A test particle of mass $m$ after
swallowed by a black hole increases the black hole mass and its
surface. This procedure signifies information carried by the test
particle lost to the increased black hole entropy or
holographic screen.

The entropic force idea expands the information and holographic
connection to flat non-relativistic space, not near the black hole.
Verlinde postulated that the entropy $\Delta S$ gained on a
holographic screen associated with a test particle of mass $m$
moving by a distance $\Delta x$ perpendicular towards the screen is
given by~\cite{Verlinde}
\begin{eqnarray}
\Delta S = 2\pi k {mc \over \hbar}\Delta x\;.
\end{eqnarray}

This equation, although a postulation, is very intuitive from the
point of view that the entropy gained is proportional to the
information loss of the test particle. The particle has a Compton
wavelength $ \lambdabar_m= {\hbar}/{mc}$ which probes information of
the particle experienced on the pass. After traveling a distance
$\Delta x$, a natural gauge of information bits is proportional to
$\Delta x/\lambdabar_m$. Several papers have discussed the
properties of the quantity $\Delta S/
k$~\cite{Kothawala,Padma,Padma2}, which is argued to be quantized in
units of $2\pi$. One therefore can write
\begin{equation}
\Delta S=   2\pi k \frac{{\Delta} x}{ \lambdabar_{m}},
\label{verlinde}
\end{equation}
Associating the information bits with the wavelength, instead of the
mass of a particle has the advantage, from quantum mechanics of
view, when one tries to expand the same idea to massless particles,
such as a photon. A massless particle moving towards a holographic
screen should also has similar behavior. We will come back to this
when we discuss entropic force on photons later.

To obtain Newton's second law, Verlinde associates temperature $T$
in the term $T{\mathrm{d}}S$ with the Unruh~\cite{Unruh} temperature
$T_u$ by
\begin{equation}
T= T_u=\frac{\hbar a}{2\pi c k},
\label{Unruh}
\end{equation}
where $a$ denotes the acceleration experienced by the test particle.

For the isolated system composed of a holographic screen and a test
particle, the tendency for an increase of the entropy of the system
gives rise of an emergent force $F$ on the test particle, i.e.,
\begin{equation}
F\Delta x=T \Delta S,
\label{FxTS}
\end{equation}
Thus one immediately gets Newton's second law
\begin{equation}
F=ma.
\label{Fma2}
\end{equation}

A system of total mass $M$ within the holographic screen is assumed
to be related to the total information bits $N$ by the equipartition
rule~\cite{Padma2}
\begin{equation}
Mc^2 = E=\frac{1}{2}k T N,
\label{E2}
\end{equation}
where $N$ is the number of quantized bits on the holographic screen
\begin{equation}
N=\frac{A}{l_P^2}=\frac{A c^3}{G \hbar}.
\end{equation}
This means that the area of the holographic screen has to be
quantized in units of the Planck area $l_P^2$. One then obtains
\begin{equation}
T=\frac{2Mc^2}{k N}=\frac{G M}{R^2}\frac{\hbar}{2\pi k c},
\label{TEE}
\end{equation}
in which $A=4\pi R^2$ is adopted. Substituting this into Eq.~(\ref{Unruh}), one arrives at the
surface acceleration
\begin{equation}
a=\frac{2\pi c k}{\hbar }T=\frac{GM}{R^2}.
\end{equation}
Thus Eq.~(\ref{Fma2}) changes into
\begin{equation}
F=G \frac{M m}{R^2},
\label{NLG}
\end{equation}
which is the famous Newton's law of gravity.

Alternatively, substituting Eq.~(\ref{TEE}) into Eq.~(\ref{FxTS})
together with the Verlinde conjecture Eq.~(\ref{verlinde}), one still arrives at
the gravity law Eq.~(\ref{NLG}), without using the Unruh temperature Eq.~(\ref{Unruh}).

Note that the sizes of the holographic screen containing a fixed total energy $E = Mc^2$ are different
when a test particle is at different locations $R$ and $r$ because the force are $GMm/R^2$ and $GMm/r^2$, respectively.
This is a situation which can be applied to the situation for a test particle on the Earth and a test particle on Mars in the solar system.
A consistent picture can be obtained by noticing that the Unruch temperatures
at these two locations are also different $T_R/T_r=r^2/R^2$, and
identifying the sizes of the holographic screens felt by the test particle
depend on the location given by $A_R = 4\pi R^2$ and $A_r = 4 \pi r^2$,
The equipartition rule still holds.
\\

\noindent {\bf Black Hole from Entropic Consideration}
\\

For a black hole, its entropy is completely determined by the black
hole surface $A_{\mathrm{BH}}$. A radial increase $\Delta R$ of a
black hole with horizon radius $R_{\mathrm H}$ causes an increase of
entropy
\begin{eqnarray}
\Delta S_{\mathrm{BH}}=\frac{ k 8\pi R_{\mathrm H}  \Delta R}{4 l_{P}^2}.
\label{b-entropy-change}
\end{eqnarray}

If the black hole horizon is the holographic screen of a test
particle of $m$ just outside and moves a distance $\Delta x$, one
would have, according to Verlinde's conjecture,
\begin{eqnarray}
\frac{2\pi  k R_{\mathrm{H}}  }{l_{\mathrm P}^2} \Delta R = \frac{
2\pi k m c}{\hbar} \Delta x\;, \label{B-holographic}
\end{eqnarray}
which leads to
\begin{eqnarray}
R_{\mathrm{H}} \Delta R = {l^2_P \Delta x \over \lambdabar_m} =
\frac{2 G m}{c^2} {\Delta x \over 2}\;.
\end{eqnarray}
Since $R_h = 2 Gm/c^2$ is the Schwarzschild radius of black hole
with mass $m$, one obtains a relation
\begin{eqnarray}
R_{\mathrm{H}} \Delta R = R_h \Delta x/2. \label{relation}
\end{eqnarray}

There is a question on how to understand this relation. In the
special case $m = M$, $R_{\mathrm H} = R_h$, one therefore obtains
$\Delta R = \Delta x/2$ to have a consistent picture. This can be
understood from the point of view that the change $\Delta x$ in
Eq.~(\ref{verlinde}) is linear in one dimension, but $\Delta R$ in
Eq.~(\ref{b-entropy-change}) causes a change for a two dimensional
surface and therefore should be a factor 2 less. For $m$ not equal
to $M$, in general, one should set the entropy change in $\Delta S$
to be in unit of $2\pi k$ after a displacement $\Delta x$, and the
entropy change in $\Delta S_\mathrm{BH}$ to be in units of $4\pi k$
after a change in radius $\Delta R$. If $\Delta S$ is quantized in
unit $2\pi k$ as suggested in Refs.~\cite{Kothawala,Padma,Padma2},
$\Delta S_\mathrm{BH}$ should quantized in units of $4\pi k$.

We can re-write the relation Eq.~(\ref{relation}) as
\begin{eqnarray}
\frac{\Delta R}{ \lambdabar_M} =  \frac{\Delta x}{2 \lambdabar_m}.
\label{relation-lambda}
\end{eqnarray}
The above equation indicates that the number of entropy in basic units recorded in $\Delta R$
by the holographic screen with two dimensional degrees of freedom
should be half of that lost by the test particle moving $\Delta x$ with only one
dimensional degree of freedom. Then we find the equivalence between
the entropy acquainted by the black hole during a horizon radial
increase $\Delta R$ and that in Verlinde's conjecture, i.e.,
Eq.~(\ref{B-holographic}).

This implies that our knowledge of black holes is consistent with
Verlinde's conjecture. One can also re-derive the black hole
properties, such as the Schwarzschild radius and the entropy
formula, from an entropic viewpoint without Newtonian mechanism and
relativity.

Our knowledge of black holes is based on classical mechanism,
relativity, and quantum theory. Reversely, one may also take
Eq.~(\ref{relation-lambda}) as a basic Ansatz, instead of Verlinde's
conjecture, to re-derive the second law of motion and the law of
gravity, based on knowledge of black holes.

At this point we should like to suggest an entropic framework with
a rule regarding entropy changes in different dimensions.
The entropy change caused by a linear displacement $\Delta l$ in units of its Compton wavelength
$\lambdabar$ is given by
\begin{equation}
\Delta S= 2\pi k D \frac{\Delta l}{\lambdabar},
\end{equation}
where $D$ is the dimensional
degree of freedom of the object under consideration. For example in our analysis
$D=1$ for a test particle and $D=2$ for a holographic screen.
If $\Delta S$ is quantized in unit $2\pi k$ as suggested
in Refs.~\cite{Kothawala,Padma,Padma2}, the entropy change for a $D$ dimensional object is then given by $2\pi k D$.

This new rule unifies Verlinde's conjecture and the entropy change
of black hole by a radial change $\Delta R$. It extends Verlinde's
conjecture to a more general case where the resultant change of
entropy is encoded in a multi-dimensional surface due to a linear
displacement.
\\

\noindent {\bf Entropic Force for Photon}
\\

So far the entropic force has been considered for massive particles.
The behaviors of the photon in gravitational environments cannot be
handled in classical mechanism as the photon mass is zero. A photon
should have similar property in carrying information just like a
massive particle should.  Eq.~(\ref{verlinde}) already provided a
hint how entropic force can be generalized to photon since a photon
with energy $E_{\gamma}$ has a Compton wavelength $\lambdabar={\hbar
c}/{E_{\gamma}}$ from quantum theory. To this end we propose that
the entropy increase on the screen for a photon moves towards a
holographic screen is
\begin{equation}
\Delta S=   2\pi k \frac{{\Delta} x}{ \lambdabar}.
\label{verlinde2}
\end{equation}

From $F\Delta x= T\Delta S$
one gets
\begin{equation}
F=\frac{2\pi k T}{\lambdabar}=\frac{2\pi k T}{\hbar c} E_{\gamma},
\end{equation}
where $T$ is the temperature of the holographic screen from Eq.~(\ref{TEE}). Thus we obtain
the gravitational force on the photon
\begin{equation}
F=\frac{GM}{R^2}\frac{E_{\gamma}}{c^2}=G \frac{M m_{\gamma}}{R^2},
\end{equation}
as if the photon has a gravitational mass $m_{\gamma}$ from the
relation $E_{\gamma}= m_{\gamma} c^2$. Such a conclusion is the same
as that from the equivalence principle of general relativity,
therefore it gives the same predictions of the gravitational
red/blue shift and bending of light by gravity. The entropic
analysis brings also new insights on entropic force for photon.
Similarly one can generalize to all massless particles.
\\

\noindent{\bf Discussion}
\\

We have studied some implications of entropic force proposed
recently by Verlinde. We find that the size of and
temperature on the holographic screen felt by a test particle depend
on the distance of the test particle away from the center of the
energy contained inside the screen.

When applied the entropic analysis to a black
hole, we find that since the entropy is completely determined by the
surface, the change of radius $\Delta R$ of a black hole, due to a
test particle moving towards the black hole by $\Delta x$ near the
black body surface, is given by $ \Delta R= {R_h \Delta x}/{2
R_{\mathrm{H}}}$, from which one has  $R_{\mathrm{H}}\Delta R= R_h
\Delta x/ 2$ or ${\Delta R}/{ \lambdabar_M} = {\Delta x}/{2
\lambdabar_m}$. Such a relation implies a consistency between
Verlinde's conjecture and the entropy acquainted by the black hole
during a radial increase. We suggested a new measure for entropy change in units of $2\pi k D$
for a $D$ dimensional object. For example with $D =1 $ and $D=2$ for a linear movement of a test particle
and a black hole surface, respectively.

When expanding the entropic
force idea to massless particles such as a photon, we find that
there is an entropic force on a photon of energy $E_\gamma$, with
$F=G M m_{\gamma}/R^2$, and therefore the photon has an effective
gravitational mass $m_\gamma = E_\gamma/c^2$ leading to blue and red
shifts, and also bend of light for a photon passing by a massive
body.

From our analysis it is clear that the holographic principle, the concept of
Unruh temperature and the equipartition rule offer natural derivations
for the second law of
motion and the gravity law in Newton's theory, the properties of
black holes, and also the gravitational effect on photons. Therefore
we may consider the entropic framework as an axiomatic system in
parallel to Newtonian mechanism, Lagrangian mechanism, or
Hamiltonian mechanism.
%principle of least action in classical mechanism.
It is a new language for the conventional knowledge of classical
mechanism. However, there are new perspectives beyond the classical
mechanism from this new framework, as have been seen by the
gravitational effect on photons revealed in this work. Therefore we
are optimistic that the entropic framework can offer us insights to
understand physics from a new way.

\begin{acknowledgments}
This work is partially supported by NSC, NCTS, NSFC (Nos.~10721063, 10975003). BQM acknowledges the support of the LHC physics focus group of NCTS and warm hospitality from
W.-Y. Pauchy Hwang during his visit of NTU.

\end{acknowledgments}

\end{document}